\documentclass[a4paper,11pt]{article}

\usepackage{jheppub} 

\usepackage[T1]{fontenc} 
\usepackage{mathrsfs}

\title{\boldmath  Revisit  spin effects induced by thermal vorticity}

\author[a,1]{Jian-Hua Gao.\note{Corresponding author.}}
\author[b]{Shi-Zheng Yang,}

\affiliation[a]{Shandong Key Laboratory of Optical Astronomy and Solar-Terrestrial Environment,
School of Space Science and Physics, Shandong University, Weihai,Shandong, 264209, China}
\affiliation[b]{Institute of Frontier and Interdisciplinary Science,
Key Laboratory of Particle Physics and Particle Irradiation (MOE), Shandong University, Qingdao, Shandong 266237, China}

\emailAdd{gao@sdu.edu.cn}
\emailAdd{yangshizheng@mail.sdu.edu.cn}

\abstract{We revisit the  spin  effects  induced by thermal vorticity by calculating them
directly from the spin-dependent distribution functions. For the  spin-1/2 particles, we give the  polarization
up to the first order of thermal vorticity  and compare it with the usual result calculated from the  spin vector.
For the spin-1 particles, we  find that all the non-diagonal elements vanish and there is no spin alignment up the first order of thermal vortcity. We present  the spin alignment at  second-order contribution from thermal vorticity.
We also find that the spin effects for both Dirac and vector particles will receive extra contribution
when the spin direction is associated with the particle's momentum.
}

\begin{document}
\maketitle
\flushbottom

\section{Introduction}
\label{sec:intro}
The spin polarization effects have been observed in heavy-ion collisions at RHIC \cite{STAR:2017ckg,Adam:2019srw,Adam:2020pti,STAR:2022fan} and LHC \cite{Acharya:2019vpe,ALICE:2021pzu} since the pioneering theoretical prediction\cite{Liang:2004ph,Liang:2004xn,Gao:2007bc}.
Relevant  reviews on spin effects in  heavy ion collisions can be find in Refs
\cite{Liang:2007ma,Wang:2017jpl,Liang:2019clf,Florkowski:2018fap,Becattini:2020ngo,Liu:2020ymh,Gao:2020vbh,Gao:2020lxh,Huang:2020dtn,Becattini:2021lfq}.
However some recent measurement results are  contradictory to the theoretical calculation such as the local longitudinal  polarization of hyperons \cite{Becattini:2017gcx} and the spin alignment of vector mesons \cite{Liang:2004xn}.  These  spin puzzles have promoted lots of  theoretical research \cite{Florkowski:2019voj,Xia:2019fjf,Becattini:2019ntv,Liu:2019krs,Liu:2021uhn,Fu:2021pok,Becattini:2021suc,Becattini:2021iol,
Yi:2021ryh,Xia:2020tyd,Gao:2021rom,Sheng:2019kmk,Sheng:2020ghv,Sheng:2022wsy,Sheng:2022ffb,Wei:2023pdf,Li:2022vmb,
Muller:2021hpe,Kumar:2023ghs,Wang:2021owk,Fang:2022ttm,Lin:2021mvw,Lin:2022tma}  on these topics. Among them, some new physical mechanisms are proposed to account for these spin puzzles,  such as the shear contribution for the local longitudinal polarization of hyperons \cite{Liu:2021uhn,Fu:2021pok,Becattini:2021suc,Becattini:2021iol}
 and strong force fields for the spin-alignment of vector mesons \cite{Sheng:2019kmk,Sheng:2020ghv,Sheng:2022wsy,Sheng:2022ffb}.
While we are  resorting  to new physical mechanisms to interpret these unexpected  results, it is also necessary to  revisit the
original theoretical methods which resulted in these discrepancies to see if we could modify and improve them in some way.  Such revisiting process would be indispensable to pin down the real physical mechanism for the spin polarization effects in heavy-ion
collisions in quantitative level.

For the spin polarization of the hyperon, the  numerical prediction  in the formalism of relativistic hydrodynamics is based on the spin Cooper-Frye formula \cite{Becattini:2013fla}
which measures the  mean spin  vector $S^\mu(k)$  by integrating  local spin  vector $S^\mu(x,k)$ over the freeze-out surface $\Sigma_\alpha$ in the heavy-ion collisions
\begin{eqnarray}
\label{S-mu-p}
S^\mu(k) &=& \frac{\int d\Sigma_\alpha k^\alpha S^\mu(x,k)  f(x,k)}
{\int d\Sigma_\alpha k^\alpha f(x,k) }
\end{eqnarray}
where $f(x,k) $ is single-particle  distribution function at the space-time point $x^\mu=(t,{\bf x})$
 with four momentum $k^\mu=(E_{\bf k},{\bf k})$. For the particle with mass $m$, the energy is given by
 $E_{\bf k}=\sqrt{m^2+{\bf k}^2}$.  At global thermodynamical equilibrium with small thermal vorticity $\varpi_{\mu\nu}$, the first order contribution for the spin vector   $S^\mu(x,k)$ is given by
\begin{eqnarray}
\label{S-mu-x-p}
S^\mu(x,k) &=&\frac{f'_F}{8mf_F} \epsilon^{\mu\nu\rho\sigma} k_\nu  \varpi_{\rho\sigma}
\end{eqnarray}
where $f_F $ is Fermi-Dirac distribution function
\begin{eqnarray}
\label{FD}
f_F=\frac{1}{e^{\beta^\mu k_\mu-\alpha}+1}
\end{eqnarray}
The  four-temperature vector $\beta^\mu  =(\beta^0, \pmb{\beta})$ is related to   fluid velocity $u^\mu$ with $u^2=1$ and   temperature $T$ by $\beta^\mu = u^\mu/T$ and $\alpha=\mu/T$ denotes the chemical potential $\mu$ scaled by temperature $T$.
The thermal vorticity is defined as
\begin{eqnarray}
\varpi_{\mu\nu}=-\frac{1}{2}\left(\partial_\mu \beta_\nu -\partial_\nu \beta_\mu\right),
\end{eqnarray}
with the components
\begin{eqnarray}
\varepsilon^{i}= \varpi^{i 0} = -\frac{1}{2}\left(\partial^i \beta^0 - \partial_t \beta^i\right),\ \ \
\omega^i = \frac{1}{2}\epsilon^{0ijk}\varpi_{jk} = -\frac{1}{2} \epsilon^{0ijk}\partial_j\beta_k,\ \ \
\end{eqnarray}
or  in 3-vector form
\begin{eqnarray}
\pmb{\varepsilon} = \frac{1}{2}\left(\pmb{\nabla} \beta^0 + \partial_t \pmb{\beta}\right), \ \ \
\pmb{\omega} = \frac{1}{2}\pmb{\nabla}\times \pmb{\beta}
\end{eqnarray}
After transforming the mean spin vector  $S^\mu=(S^0,{\bf S})$ in Eq.(\ref{S-mu-p}) into the rest frame of the particle with momentum
${\bf k}$
\begin{eqnarray}
S^{* \mu} =  \left(0 ,  {\bf S}^* \right),\ \ \
{\bf S}^* = {\bf S } -  \frac{({\bf S}\cdot {\bf k}) {\bf k}}{E_{\bf k}(E_{\bf k}+m)},
\end{eqnarray}
the final polarization $P$ along some quantization direction ${\bf n}_3$ is given by
\begin{eqnarray}
P={\bf n}_3\cdot {\bf S}^* &=& {\bf n}_3\cdot {\bf S } -  \frac{({\bf S}\cdot {\bf k})({\bf n}_3\cdot {\bf k})}{E_{\bf k}(E_{\bf k}+m)}
\end{eqnarray}
With the specific expression (\ref{S-mu-x-p}) for local spin vector, the 3-vector form reads
\begin{eqnarray}
S^0(x,k)=\frac{f'_F}{4m f_F} \pmb{\omega}\cdot {\bf k},\ \ \
{\bf S}(x,k)=-\frac{f'_F}{4m f_F}\left( E_{\bf k}\pmb{\omega} - \pmb{\varepsilon}\times {\bf k}\right)
\end{eqnarray}
Then the local polarization is given by
\begin{eqnarray}
\label{P-x-k-0}
P(x,k)= -\frac{ f'_F}{2f_F}\cdot\frac{E_{\bf k}}{2m}
\left[\pmb{\omega}-\frac{(\pmb{\omega}\cdot {\bf k}){\bf k}}{E_{\bf k} (E_{\bf k} +m)}
 - \frac{\pmb{\varepsilon}\times {\bf k}}{E_{\bf k}}\right]\cdot{\bf n}_3
\end{eqnarray}

As we all know,  the polarization $P$ can be obtained directly from the particle distribution
$f_{rs}(x,k)$ with  index  $r,s=\pm 1$ (sometimes $r,s=\pm$ for brevity) corresponding to the spin $\pm 1/2$ along the spin quantization direction ${\bf n}_3$
\begin{eqnarray}
P(x,k)=\frac{f_{+,+}(x,k)-f_{-,-}(x,k)}{f_{+,+}(x,k)+f_{-,-}(x,k)}.
\end{eqnarray}
Hence we can also calculate  the final polarization in heavy-ion collisions with
\begin{eqnarray}
\label{P-p}
P(k) &=& \frac{\int d\Sigma_\alpha k^\alpha P(x,k)  f(x,k)}
{\int d\Sigma_\alpha k^\alpha f(x,k) }
\end{eqnarray}
where $f(x,k)\equiv f_{++}(x,k)+f_{--}(x,k)$ means the sum of spin up and spin down along the direction ${\bf n}_3$.
We will demonstrate that the polarization from (\ref{P-p}) is  different from (\ref{S-mu-p}).

For the spin polarization of the vector meson,  theoretical predictions  focuses on the spin alignment
and there is no  similar formula as Eq.(\ref{S-mu-p}) for the vector meson yet.
Most prediction relies on the quark coalescence model.
In the formalism of relativistic hydrodynamics, the measured spin density matrix can be calculated
from the particle distribution function $f_{rs}(x,k)$ with spin index $r,s = 0,\pm 1$
\begin{eqnarray}
\label{rho-sr-p}
\rho_{rs}(k) &=& \frac{\int d\Sigma_\alpha k^\alpha f_{rs}(x,k) }
{\int d\Sigma_\alpha k^\alpha f(x,k) }
\end{eqnarray}
where $f(x,k)\equiv  f_{11}(x,k)+  f_{00}(x,k) + f_{-1-1}(x,k)$ means the sum of all the diagonal components along the direction ${\bf n}_3$.
We will derive the specific expression for this density matrix by calculating the particle distribution $f_{rs}(x,k)$. We will show
that the spin alignment receives only  second-order contribution from accelaration or vorticity while some non-diagonal elements in spin density matrix can receive  first order contribution

We note that the particle distributions $f_{rs}(x,k)$ with spin will be the crucial elements in the formulae (\ref{P-p}) and (\ref{rho-sr-p}) rather than the spin  vector in (\ref{S-mu-p}) and
(\ref{S-mu-x-p}). Since we only revisit the  spin polarization by thermal vorticity in this work, we will calculate the particle distributions with spin for free particle in global equilibrium with thermal vorticity and assume these results will still dominate in local equilibrium. Recently the exact equilibrium distributions with thermal vorticity have been obtained by analytical continuation in Refs\cite{Becattini:2020qol,Palermo:2021hlf,Palermo:2023cup}, in our present work we will calculate these distribution functions in a more direct and usual way and expand them specifically in terms of vorticity and acceleration  in first or second order.

We first calculate the particle distributions for scalar field in global equilibrium with thermal vorticity in Sec.~\ref{sec:scalar}.
 Then we will  deal with the particle distributions for Dirac fields in Sec.~\ref{sec:dirac} and vector field in Sec.~\ref{sec:vector}.
 A summary of our results is listed in Sec.~\ref{sec:summary}.
In this work, we use  the metric $g^{\mu\nu}=\mathrm{diag}(1,-1,-1,-1)$ and Levi-Civita tensor $\epsilon^{0123}=1$.

\section{Scalar field}
\label{sec:scalar}

Let us first review some well-known results for the scalar field. The Lagrange density of charged scalar field reads
\begin{eqnarray}
\mathcal{L}=\partial^\mu\phi^\dagger \partial_\mu\phi -m^2\phi^\dagger\phi.
\end{eqnarray}
The Euler-Lagrangian  equation is just Klein-Gordon equation
\begin{eqnarray}
\partial^\mu \partial_\mu\phi + m^2\phi = 0,\ \ \ \partial^\mu \partial_\mu\phi^\dagger + m^2\phi^\dagger = 0
\end{eqnarray}
From Noether' theorem,  we have the conserved charge current
\begin{eqnarray}
\label{j-KG}
j^{\mu} &=& i\left( \partial^\mu\phi^\dagger\phi - \phi^\dagger \partial^\mu\phi\right),
\end{eqnarray}
 the canonical energy-momentum tensor
\begin{eqnarray}
\label{T-KG}
T^{\mu\nu} &=&\partial^\mu\phi^\dagger \partial^\nu\phi + \partial^\nu\phi^\dagger \partial^\mu\phi
- g^{\mu\nu}\left(\partial_\alpha\phi^\dagger \partial^\alpha\phi -m^2\phi^\dagger\phi\right),
\end{eqnarray}
and the angular momentum density
\begin{eqnarray}
\label{M-KG}
\mathcal{M}^{\mu\alpha\beta}
&=&  x^\alpha T^{\mu\beta} - x^\beta T^{\mu\alpha}
\end{eqnarray}
We can expand Klein-Gordon field in terms of annihilation operator and creation operator
\begin{eqnarray}
\label{KG-Fourier}
\phi(x) &=& \int \frac{d^3 {\bf p}}{(2\pi)^3}\frac{1}{\sqrt{2E_{\bf p}}}
 \left[a_{\bf p}  e^{-ip\cdot x} + b_{\bf p}^\dagger e^{ip\cdot x} \right],\nonumber\\
\phi^\dagger(x) &=& \int \frac{d^3 {\bf p}}{(2\pi)^3}\frac{1}{\sqrt{2E_{\bf p}}}
 \left[b_{\bf p} e^{-ip\cdot x} + a_{\bf p}^\dagger e^{ip\cdot x} \right],
\end{eqnarray}
where $p^\mu = (p^0,{\bf p})$ with $p^0=E_{\bf p}$ and the creation and annihilation operators obey the commutation rules
\begin{eqnarray}
\left[a_{\bf p} , a_{\bar{\bf p}}^\dagger\right]= \left[b_{\bf p} , b_{\bar{\bf p}}^\dagger\right]
=(2\pi)^3 \delta^{(3)}({\bf p}-\bar{\bf p})
\end{eqnarray}
Inserting the expression (\ref{KG-Fourier}) into Eqs.(\ref{j-KG}) and \ref{T-KG}) and integrating over the whole space, we obtain the conserved charge and energy-momentum
\begin{eqnarray}
\label{Q-KG}
Q &=& \int d^3 {\bf x} \,j^{0} = \int \frac{d^3 {\bf p}}{(2\pi)^3}
  \left( a_{\bf p}^{\dagger} a_{\bf p}
-  b_{\bf p} b_{\bf p}^{\dagger} \right),\\
\label{Pmu-KG}
P^\mu &=&  \int d^3 {\bf x} \, T^{0\mu} = \int \frac{d^3 {\bf p}}{(2\pi)^3}
 p^\nu  \left( a_{\bf p}^{\dagger} a_{\bf p}
+  b_{\bf p} b_{\bf p}^{\dagger} \right).
\end{eqnarray}
The angular-momentum tensor is defined  by
\begin{eqnarray}
J^{\mu\nu}&=&  \int d^3 {\bf x} \, \mathcal{M}^{0\mu\nu},
\end{eqnarray}
with the components
\begin{eqnarray}
\label{K-i-KG}
K^{i}&=&  J^{0i} = \int \frac{d^3 {\bf p}}{(2\pi)^3}
\left[\sqrt{E_{\bf p}}a_{\bf p}^{\dagger} i\partial^i_p \left( \sqrt{E_{\bf p}} a_{\bf p}\right)
- \sqrt{E_{\bf p}} b_{\bf p} i\partial^i_p \left( \sqrt{E_{\bf p}} b_{\bf p}^{\dagger}\right)\right],\\
\label{J-i-KG}
J^{i}&=&\frac{1}{2} \epsilon^{0ijk} J_{jk} = \epsilon^{0ijk} \int \frac{d^3 {\bf p}}{(2\pi)^3}
p^k\left[a_{\bf p}^{\dagger}i\partial_j^p a_{\bf p}
-  b_{\bf p} i\partial_j^p b_{\bf p}^{\dagger}\right]
\end{eqnarray}
or in 3-vector form
\begin{eqnarray}
{\bf K}&=&- \int \frac{d^3{\bf p}}{(2\pi)^3}
\left[ a_{\bf p}^{\dagger} \sqrt{E_{\bf p}}\pmb{\nabla}_p  \left( \sqrt{E_{\bf p}}a_{\bf p}\right)
-  b_{\bf p}\sqrt{E_{\bf p}}\pmb{\nabla}_p\left( \sqrt{E_{\bf p}} b_{\bf p}^{\dagger}\right)\right],\\
{\bf J} &=&-\int \frac{d^3{\bf p}}{(2\pi)^3}
\left[a_{\bf p}^{\dagger} \left({\bf p}\times i\pmb{\nabla}_p\right) a_{\bf p}
-  b_{\bf p} \left({\bf p}\times i\pmb{\nabla}_p\right) b_{\bf p}^{\dagger}\frac{}{}\right]
\end{eqnarray}
where
\begin{eqnarray}
\partial_p^i = \frac{\partial}{\partial p_i} = - \frac{\partial}{\partial p^i}, \ \ \ \ \
\pmb{\nabla}_p  = \left(\frac{\partial}{\partial p^1},\frac{\partial}{\partial p^2}, \frac{\partial}{\partial p^3} \right)
\end{eqnarray}

Now let us turn to the particle distribution function which  is defined  by \cite{DeGroot:1980dk}
\begin{eqnarray}
\label{f-KG}
f(x,k)&\equiv&
\frac{1}{(2\pi)^3}\int d^3 {\bf q} e^{-i(E_{{\bf k}+{\bf q}/2}-E_{{\bf k}-{\bf q}/2})t + i{\bf q}\cdot{\bf x}}
\langle a^{\dagger}_{{\bf k}-{\bf q}/{2}} a_{ {\bf k}+{\bf q}/{2}}\rangle
\end{eqnarray}
where $\langle Q \rangle =  \textrm{Tr}\left( \rho Q\right)$ denotes ensemble average of some operator $Q$ with  density matrix $\rho$. Here we  choose
the density operator in global equilibrium \cite{Becattini:2015nva}
\begin{eqnarray}
\label{rho}
\rho&=&\frac{1}{Z}\textrm{exp}\left(-b_\mu P^\mu  +\alpha  Q + \frac{1}{2}\omega_{\mu\nu}J^{\mu\nu} \right),\ \ \
\end{eqnarray}
with $Z$ the partition function, $b_\mu$ constant timelike vector, $\alpha$ constant scalar and $\omega_{\mu\nu}$ constant antisymmetric tensor.  The operators  $P^\mu $, $Q$ and $J^{\mu\nu}$ are the energy-momentum, charge and angular momentum tensor, respectively.  The contribution from the angular momentum tensor can rewritten as
\begin{eqnarray}
\frac{1}{2}\omega_{\mu\nu}J^{\mu\nu}
&=& {\pmb{\varepsilon}}\cdot{\bf K} + {\pmb{\omega}}\cdot{\bf J}.
\end{eqnarray}
The density operator (\ref{rho}) is a general result and valid for any field. For the free charged scalar field, $P^\mu $, $Q$ and $J^{\mu\nu}$ will take the specific forms in
Eqs.(\ref{Q-KG}),(\ref{Pmu-KG}), (\ref{K-i-KG}), and (\ref{J-i-KG}).

In  order to get the distribution function $f(x,k)$, we first calculate
\begin{eqnarray}
\langle a^{\dagger}_{\bf p} a_{ \bar{\bf p}}\rangle
=\frac{1}{Z} \textrm{Tr}\left[\textrm{exp}\left(-b_\mu P^\mu  +\alpha  Q + \frac{1}{2}\omega_{\mu\nu}J^{\mu\nu} \right) a^{\dagger}_{\bf p} a_{ \bar{\bf p}} \right].
\end{eqnarray}
From the Eqs.(\ref{Q-KG}),(\ref{Pmu-KG}), (\ref{K-i-KG}), and (\ref{J-i-KG}), the following commutation relations hold
\begin{eqnarray}
\left[Q, a_{\bf p}^{\dagger}\right] =  a_{{\bf p}}^{\dagger},\ \ \
\left[b_\mu P^\mu, a_{\bf p}^{\dagger}\right] = b_\mu p^\mu a_{{\bf p}}^{\dagger},\ \ \
\left[{\pmb{\varepsilon}}\cdot{\bf K} + {\pmb{\omega}}\cdot{\bf J}, a_{\bf p}^{\dagger}\right]
= \Lambda_{\bf p}{a_{\bf p}}^\dagger.
\end{eqnarray}
In the last equation above,  $\Lambda_{\bf p}$ denotes an operator defined by
\begin{eqnarray}
\label{Lambda}
\Lambda_{\bf p} &=&E_{\bf p} \pmb{\mathcal{E}}_{\bf p} \cdot i\pmb{\nabla}_{p}
 + \frac{i}{2E_{\bf p}}\pmb{\varepsilon}\cdot {\bf p},
\end{eqnarray}
where we defined an effective acceleration vector $\pmb{\mathcal{E}}_{\bf p} $
\begin{eqnarray}
\pmb{\mathcal{E}}_{\bf p} &=& \pmb{\varepsilon} +\frac{1}{E_{\bf p}} \pmb{\omega}\times {\bf p}.
\end{eqnarray}
Using the Baker-Hausdorff formula
\begin{eqnarray*}
e^{- A} B e^{ A}
=B +\frac{(-1)^1}{1!} \left[A,B\right]  + \frac{(-1)^2}{2!}\left[A,\left[A,B\right]\right] +\frac{(-1)^3}{3!}\left[A,\left[A, \left[A,B\right]\right]\right]
+\cdots
\end{eqnarray*}
we have the identity
\begin{eqnarray*}
\textrm{exp}\left(-b_\mu P^\mu  +\alpha  Q + \frac{1}{2}\omega_{\mu\nu}J^{\mu\nu} \right) a_{\bf p}^{\dagger}
\textrm{exp}\left(b_\mu P^\mu  -\alpha  Q - \frac{1}{2}\omega_{\mu\nu}J^{\mu\nu} \right)
=e^{-b_\mu p^\mu + \alpha + \Lambda_{\bf p}}a_{\bf p}^{\dagger}.
\end{eqnarray*}
which leads to
\begin{eqnarray}
\label{a-p-pbar}
\langle a^{\dagger}_{\bf p} a_{ \bar{\bf p}}\rangle
=e^{-b_\mu p^\mu + \alpha + \Lambda_{\bf p}}\langle  a_{ \bar{\bf p}} a^{\dagger}_{\bf p}\rangle
\end{eqnarray}
From the commutation relation for Klein-Gorden field, we also have
\begin{eqnarray}
\langle  a_{ \bar{\bf p}} a^{\dagger}_{\bf p} \rangle &=&(2\pi)^3\delta\left({\bf p}-\bar{\bf p} \right)
+ \langle a^{\dagger}_{\bf p} a_{ \bar{\bf p}}\rangle.
\end{eqnarray}
Substituting it into Eq.(\ref{a-p-pbar}) leads to
\begin{eqnarray}
\langle a^{\dagger}_{\bf p} a_{ \bar{\bf p}}\rangle
=(2\pi)^3\left[\frac{1}{e^{b_\mu p^\mu - \alpha - \Lambda_{\bf p}}-1}\right]\delta\left({\bf p}-\bar{\bf p} \right)
\end{eqnarray}
Replacing ${\bf p}$ and $\bar{\bf p}$ with ${{\bf k}-{\bf q}/{2}}$ and ${{\bf k}-{\bf q}/{2}}$, respectively, and inserting the above expression into Eq.(\ref{f-KG}),  we obtain the distribution function
\begin{eqnarray}
\label{f-KG-1}
f(x,k)
=\int d^3 {\bf q} e^{-i(E_{{\bf k}+{\bf q}/2}-E_{{\bf k}-{\bf q}/2})t+i{\bf q}\cdot{\bf x}}
\frac{1}{e^{b_0 E_{{\bf k}-{\bf q}/2} - {\bf b}\cdot({\bf k}-{\bf q}/2) - \alpha - \Lambda_{{\bf k}-{\bf q}/2}}-1}
\delta\left({\bf q}\right)
\end{eqnarray}
where the operator $\Lambda_{{\bf k}-{\bf q}/2}$ reads
\begin{eqnarray}
\Lambda_{{\bf k}-{\bf q}/2} &=&E_{{\bf k}-{\bf q}/2} \pmb{\mathcal{E}}_{{\bf k}-{\bf q}/2}
\cdot i\left(\frac{1}{2}\pmb{\nabla}_{k} - \pmb{\nabla}_{q}\right)
 + \frac{i}{2E_{{\bf k}-{\bf q}/2}}\pmb{\varepsilon}\cdot \left( {\bf k}-\frac{\bf q}{2}\right).
\end{eqnarray}
We assume that   ${\pmb{\varepsilon}}$ and $ {\pmb{\omega}}$ are both small variables and expand the distribution function as the Taylor series of these variables.
We need the following expansion
\begin{eqnarray}
\label{Expansion-BE}
\frac{1}{e^{X+Y} - 1}
&=&f_{B}(X)+f'_{B}(X)Y - \frac{1}{2}f''_{B}(X)C  +\frac{1}{2}f''_{B}(X)Y^2\nonumber\\
& & - \frac{1}{6}f'''_B(X)YC - \frac{1}{3}f'''_B(X)CY +\frac{1}{8}f''''_B(X)C^2+\cdots\hspace{1cm}
\end{eqnarray}
where $C=[X,Y]$ and $f_B(X) $ is Bose-Einstein distribution function $1/(e^{X}-1)$. This expansion is valid up to the second order of $Y$. In our current case, we can identify $X$ and $Y$ as
\begin{eqnarray}
X&=&b_0 E_{{\bf k}-{\bf q}/2} - {\bf b}\cdot({\bf k}-{\bf q}/2) - \alpha,\ \ \ \
Y=-\Lambda_{{\bf k}-{\bf q}/2} \\
C&=&[X,Y]=i\left[\left(b_0 \pmb{\varepsilon}+ \pmb{\omega}\times {\bf b}\right)
\cdot \left( {\bf k}-\frac{\bf q}{2}\right) - ({\bf b}\cdot\pmb{\varepsilon})E_{{\bf k}-{\bf q}/2}\right]
\end{eqnarray}
Now we can calculate the distribution functions order by order after integrating over the momentum ${\bf q}$. The zeroth-order result is trivial and just the Bose-Einstein distribution function
\begin{eqnarray}
f^{(0)}(x,k)
&=&f_B(b\cdot k -\alpha)
\end{eqnarray}
The first-order result is also  simple
\begin{eqnarray}
f^{(1)}(x,k)
&=& f'_B(b\cdot k -\alpha)
\left[E_{\bf k}\pmb{\varepsilon}\cdot{\bf x} - {\bf k}\cdot (\pmb{\varepsilon}t + \pmb{\omega}\times{\bf x})\right]
\end{eqnarray}
The second-order result is a little complicated
\begin{eqnarray}
\label{f-2-KG}
f^{(2)}(x,k)
&=&\frac{1}{2} f''_B(b\cdot k -\alpha)
\left[E_{\bf k}\pmb{\varepsilon}\cdot{\bf x} - {\bf k}\cdot (\pmb{\varepsilon}t + \pmb{\omega}\times{\bf x})\right]^2\nonumber\\
& & + \frac{1}{8}f''_B(b\cdot k -\alpha)
\left[\frac{({\bf k}\cdot \pmb{\varepsilon})^2}{E_{\bf k}^2}
+ \frac{2{\bf k}\cdot (\pmb{\omega}\times\pmb{\varepsilon} )}{E_{\bf k}} -2\pmb{\omega}^2 \right]\nonumber\\
& & + \frac{1}{12}f'''_B(b\cdot k -\alpha) \left(E_{\bf k}\pmb{\varepsilon}+\pmb{\omega}\times{\bf k} \right)
\cdot \left(b_0 \pmb{\varepsilon}+ \pmb{\omega}\times {\bf b}\right)
\end{eqnarray}
We note that there exist some terms which depend on the time $t$ and space coordinates ${\bf x}$. When $t$ or ${\bf x}$ is large,
our expansion will be broken. Actually we can absorb these terms  into  the vector $b^\mu$  and obtain a new vector $\beta^\mu$ which can be regarded as
the zeroth order contribution
\begin{eqnarray}
\beta^0 = b^0 + \pmb{\varepsilon}\cdot{\bf x},\ \ \ \
\pmb{\beta} = {\bf b} +  \pmb{\varepsilon} t + \pmb{\omega}\times {\bf x}
\end{eqnarray}
Summing over the particle distributions up to the second order, we final obtain
\begin{eqnarray}
\label{f-scalar}
f(x,k)
&=& f_B(\beta\cdot k -\alpha)
 + \frac{1}{8}f''_B(\beta\cdot k -\alpha)
\left[\frac{({\bf k}\cdot \pmb{\varepsilon})^2}{E_{\bf k}^2}
+ \frac{2{\bf k}\cdot (\pmb{\omega}\times\pmb{\varepsilon} )}{E_{\bf k}} -2\pmb{\omega}^2 \right]\nonumber\\
& & + \frac{1}{12}f'''_B(\beta\cdot k -\alpha) \left(E_{\bf k}\pmb{\varepsilon}+\pmb{\omega}\times{\bf k} \right)
\cdot \left(b^0 \pmb{\varepsilon}+ \pmb{\omega}\times {\bf b}\right)
\end{eqnarray}
If we identify $\beta^\mu$ as the inverse temperature vector $u^\mu/T$, we arrive at the well-known conclusion that  the spin chemical potential $\omega^{\mu\nu}$ is equal to the thermal vorticity $\varpi^{\mu\nu}$ at global equilibrium.

\section{Dirac field}
\label{sec:dirac}

Now let us consider the Dirac fermions with spin-1/2. The Lagrangian for the free Dirac field is given by
\begin{eqnarray}
\mathcal{L}=\bar\psi(i\gamma^\mu \partial_\mu -m)\psi.
\end{eqnarray}
from which we can get the Dirac equations
\begin{eqnarray}
i\gamma^\mu \partial_\mu \psi(x)-m \psi(x) =0,\ \ \ i\partial_\mu \bar\psi(x)\gamma^\mu +m\bar\psi(x) =0,
\end{eqnarray}
the electric currents
\begin{eqnarray}
j^{\mu} &=& \bar \psi \gamma^\mu \psi
\end{eqnarray}
the canonical energy-momentum tensor
\begin{eqnarray}
T^{\mu\nu} &=& \frac{i}{2}\bar\psi \gamma^\mu \left[ \overrightarrow{\partial}^\nu   - \overleftarrow{\partial}^\nu   \right]\psi
\end{eqnarray}
and the  angular momentum tensor density
\begin{eqnarray}
\mathcal{M}^{\mu,\alpha\beta} &=& x^\alpha T^{\mu\beta} - x^\beta T^{\mu\alpha}
 + \frac{1}{2}\bar\psi \left\{\gamma^\mu, \frac{\sigma^{\alpha\beta}}{2}\right\}\psi\nonumber\\
 &=& x^\alpha T^{\mu\beta} - x^\beta T^{\mu\alpha}
 - \frac{1}{2} \epsilon^{\mu\alpha\beta\nu} \bar\psi  \gamma_\nu \gamma_5 \psi.
\end{eqnarray}
Then the  charge, energy-momentum , and angular momentum tensor read, respectively,
\begin{eqnarray}
\label{Q-P-J}
Q= \int d^3 {\bf x} \, j^{0},\ \ \
P^\mu = \int d^3 {\bf x} \, T^{0\mu},\ \ \
J^{\mu\nu} =  \int d^3 {\bf x} \, \mathcal{M}^{0\mu\nu},\ \ \
\end{eqnarray}
Expand the free Dirac field in terms of  annihilation operator and creation operator
\begin{eqnarray}
\label{psi-psibar}
\psi(x) &=& \int \frac{d^3 {\bf p}}{(2\pi)^3}\frac{1}{\sqrt{2E_{\bf p}}}\sum_s
 \left[a^s_{\bf p} u^s(p) e^{-ip\cdot x} + {b^s_{\bf p}}^\dagger v^s(p) e^{ip\cdot x} \right];\nonumber\\
\bar\psi(x) &=& \int \frac{d^3 {\bf p}}{(2\pi)^3}\frac{1}{\sqrt{2E_{\bf p}}}\sum_s
 \left[b^s_{\bf p} \bar v^s(p) e^{-ip\cdot x} + {a^s_{\bf p}}^\dagger \bar u^s(p) e^{ip\cdot x} \right].
\end{eqnarray}
where  the creation and annihilation operators obey the anticommutation rules
\begin{eqnarray}
\left\{a^s_{\bf p} , {a^r_{\bar{\bf p}}}^\dagger\right\}= \left\{b^s_{\bf p} , {b^r_{\bar{\bf p}}}^\dagger\right\}
=(2\pi)^3 \delta^{(3)}({\bf p}-\bar{\bf p})\delta^{sr}
\end{eqnarray}
and the  index $s,r=\pm 1$ denotes the spin $\pm 1/2$.
Substituting the expansion (\ref{psi-psibar}) into the conserved charges (\ref{Q-P-J}), we have
\begin{eqnarray}
Q&=&\int \frac{d^3 {\bf p}}{(2\pi)^3}\sum_s
 \left[  {a^s_{\bf p}}^\dagger a^s_{\bf p}+ b^s_{\bf p} {b^s_{\bf p}}^\dagger
  \right],\\
P^\mu &=& \int \frac{d^3 {\bf p}}{(2\pi)^3}\sum_s
 \left[p^\mu  {a^s_{\bf p}}^\dagger a^s_{\bf p}- p^\mu b^s_{\bf p} {b^s_{\bf p}}^\dagger
  \right]
\end{eqnarray}
and the angular momentum tensor with the components as defined in Eqs.(\ref{K-i-KG}) and (\ref{J-i-KG})
\begin{eqnarray}
K^{i}&=& \int \frac{d^3 {\bf p}}{(2\pi)^3}\frac{1}{2}
\sum_s \sum_r \left\{{ v^{s\dagger}({\bf p})} b^s_{\bf p} ( i \partial^i_p ) \left[ {b^r_{\bf p}}^\dagger  v^r({\bf p}) \right]\right.\nonumber\\
& &\left.\hspace{3.3cm}+{u^{s\dagger}({\bf p})} {a^s_{\bf p}}^\dagger ( i \partial^i_p) \left[  a^r_{\bf p}   u^r({\bf p})\right]
\frac{}{} \right\},\\
J^i &=&     \int \frac{d^3 {\bf p}}{(2\pi)^3} \frac{1}{ 2 E_{\bf p} }\sum_s \sum_r
\left\{  v^{s\dagger}({\bf p}) b^s_{\bf p} \left[\epsilon^{ijk} p^j  (i\partial^k_p) +\frac{1}{2}\Sigma^i \right] \left[{b^{r\dagger}_{\bf p}}   v^r({\bf p})\right]\right.\nonumber\\
& &\left.\hspace{3.5cm} + u^{s\dagger}({\bf p}) {a^{s\dagger}_{\bf p}} \left[\epsilon^{ijk}  p^j ( i\partial^k_p) +\frac{1}{2}\Sigma^{i} \right] \left[ a^r_{\bf p}   u^r({\bf p})\right]\right\}
\end{eqnarray}
or in 3-vector form
\begin{eqnarray}
{\bf K}&=& \int \frac{d^3 {\bf p}}{(2\pi)^3}\frac{1}{2}
\sum_s \sum_r \left\{{ v^{s\dagger}({\bf p})} b^s_{\bf p} (- i\pmb{ \nabla}_p ) \left[ {b^r_{\bf p}}^\dagger  v^r({\bf p}) \right]\right.\nonumber\\
& &\left.\hspace{3.5cm}+{u^{s\dagger}({\bf p})} {a^s_{\bf p}}^\dagger ( - i\pmb{ \nabla}_p)
\left[  a^r_{\bf p}   u^r({\bf p})\right]\frac{}{} \right\},\\
{\bf J} &=&     \int \frac{d^3 {\bf p}}{(2\pi)^3} \frac{1}{ 2 E_{\bf p} }\sum_s \sum_r
\left\{  v^{s\dagger}({\bf p}) b^s_{\bf p} \left[- {\bf p}\times (i\pmb{\nabla}_p) +\frac{1}{2}\pmb{\Sigma} \right] \left[{b^{r\dagger}_{\bf p}}   v^r({\bf p})\right]\right.\nonumber\\
& &\left.\hspace{4cm} + u^{s\dagger}({\bf p}) {a^{s\dagger}_{\bf p}}
\left[- {\bf p}\times (i\pmb{\nabla}_p) +\frac{1}{2}\pmb{\Sigma} \right]
 \left[ a^r_{\bf p}   u^r({\bf p})\right]\right\}
\end{eqnarray}
where ${\pmb \Sigma}$ is defined from the Pauli matrix $\pmb{ \sigma}$
\begin{eqnarray}
 {\pmb \Sigma} &=& \left(
                     \begin{array}{cc}
                       {\pmb \sigma} & 0 \\
                       0 & {\pmb \sigma} \\
                     \end{array}
                   \right)
\end{eqnarray}
Then we can obtain the commutation relations
\begin{eqnarray}
\left[Q, a_{\bf p}^{s\dagger}\right] =  a_{{\bf p}}^{s\dagger},\ \ \
\left[b_\mu P^\mu, a_{\bf p}^{s\dagger}\right] = b_\mu p^\mu a_{{\bf p}}^{s\dagger},\ \ \
\left[{\pmb{\varepsilon}}\cdot{\bf K} + {\pmb{\omega}}\cdot{\bf J}, a_{\bf p}^{s\dagger}\right]
=\sum_r \Lambda_{\bf p}^{sr}{a_{\bf p}}^{r\dagger}
\end{eqnarray}
where the operator $\Lambda_{\bf p}^{sr}$ with spin index is given by
\begin{eqnarray}
\label{Lambda-sr-dirac}
\Lambda_{\bf p}^{sr}
&=&\frac{1}{2E_{\bf p}}{u^{r\dagger}({\bf p})}  u^s({\bf p})
(E_{\bf p}\pmb{\varepsilon} - {\bf p}\times \pmb{\omega})\cdot i {\pmb{ \nabla}}_p
+ \frac{1}{2E_{\bf p}}{u^{r\dagger}({\bf p})}
\left(\frac{1}{2}\pmb{\omega}\cdot{\pmb{\Sigma}}\right)  u^s({\bf p}) \nonumber\\
& &+\frac{1}{2E_{\bf p}}(E_{\bf p}\pmb{\varepsilon} - {\bf p}\times \pmb{\omega})
\cdot \left[i {\pmb{ \nabla}}_p {u^{r\dagger}({\bf p})} \right] u^s({\bf p})
\end{eqnarray}
Following the same route to arrive at Eq.(\ref{a-p-pbar}) in the scalar field, we can have
\begin{eqnarray}
\langle a^{s\dagger}_{\bf p} a^{r}_{ \bar{\bf p}}\rangle
=\left(e^{-b_\mu p^\mu + \alpha + \Lambda}\right)^{s s'}\langle  a^{r}_{ \bar{\bf p}} a^{s'\dagger}_{\bf p}\rangle
\end{eqnarray}
Thanks to the anticommutation relation for fermion field, we also have
\begin{eqnarray}
\langle  a^r_{ \bar{\bf p}} a^{s\dagger}_{\bf p} \rangle &=&(2\pi)^3\delta\left({\bf p}-\bar{\bf p} \right)\delta^{sr}
- \langle a^{s\dagger}_{\bf p} a^r_{ \bar{\bf p}}\rangle
\end{eqnarray}
which leads to
\begin{eqnarray}
\langle a^{s\dagger}_{\bf p} a^{r}_{ \bar{\bf p}}\rangle
=(2\pi)^3\left[\frac{1}{e^{b_\mu p^\mu - \alpha - \Lambda}+1}\right]^{sr}\delta\left({\bf p}-\bar{\bf p} \right)
\end{eqnarray}
It follow that the distribution function is given by
\begin{eqnarray}
\label{f-sr-Dirac}
& & f_{rs}(x,k)\equiv
\frac{1}{(2\pi)^3}\int d^3 {\bf q} e^{-i(E_{{\bf k}+{\bf q}/2}-E_{{\bf k}-{\bf q}/2})t+i{\bf q}\cdot{\bf x}}
\langle a^{s\dagger}_{{\bf k}-{\bf q}/{2}} a^r_{ {\bf k}+{\bf q}/{2}}\rangle_0\nonumber\\
&=&\int d^3 {\bf q} e^{-i(E_{{\bf k}+{\bf q}/2}-E_{{\bf k}-{\bf q}/2})t+i{\bf q}\cdot{\bf x}}
\left[\frac{1}{e^{b_0 E_{{\bf k}-{\bf q}/2} - {\bf b}\cdot({\bf k}-{\bf q}/2) - \alpha
- \Lambda_{{\bf k}-{\bf q}/2}}+1}\right]^{sr}
\delta\left({\bf q}\right)
\end{eqnarray}
Further calculation on $\Lambda_{\bf p}$ or $\Lambda_{{\bf k}-{\bf q}/2}$ need specific expressions for the four-component Dirac spinors $u^s({\bf p})$
\begin{eqnarray}
u^s({\bf p})=  \left(
           \begin{array}{c}
             \sqrt{p\cdot\sigma} \xi^s\\
             \sqrt{p\cdot \bar\sigma}\xi^s \\
           \end{array}
         \right),
\end{eqnarray}
where $\sigma^\mu = (1, \pmb{\sigma})$ and $\bar\sigma^\mu = (1, -\pmb{\sigma})$ and the two-component spinor $\xi^s$ is chosen as
\begin{eqnarray}
\xi^{+1}=\left(\begin{array}{c}
          \cos\frac{\vartheta}{2}e^{-i\varphi/2} \\
          \sin\frac{\vartheta}{2}e^{i\varphi/2}
        \end{array}\right),\ \ \ \
\xi^{-1}=\left(\begin{array}{c}
          -\sin\frac{\vartheta}{2}e^{-i\varphi/2} \\
          \cos\frac{\vartheta}{2}e^{i\varphi/2}
        \end{array}\right)
\end{eqnarray}
or in a unified form
\begin{eqnarray}
\xi^{s}=(- i)^{\frac{1-s}{2}}\left(\begin{array}{c}
          \cos\frac{s\vartheta+(1-s)\pi/2}{2}e^{-i\frac{\varphi+(1-s)\pi/2}{2}} \\
          \sin\frac{s\vartheta+(1-s)\pi/2}{2} e^{i\frac{\varphi+(1-s)\pi/2}{2}}
        \end{array}\right)
\end{eqnarray}
They are  the  eigenstates of the spin operator along the direction
${\bf n}_3=(\sin\vartheta \cos\varphi, \sin\vartheta \sin \varphi, \cos\vartheta)$:
\begin{eqnarray}
\pmb{\sigma}\cdot{\bf n}_3 \xi^{s}= s \xi^{s}
\end{eqnarray}
It is easy to verify the following relations
\begin{eqnarray}
\xi^{s}\xi^{r\dagger}
&=&\frac{1}{2}\lambda^{sr}\cdot \bar\sigma,\ \ \
\sqrt{p\cdot\sigma} =\varrho \cdot \sigma,\ \ \ \sqrt{p\cdot\bar\sigma} =\varrho \cdot \bar\sigma
\end{eqnarray}
where we have defined the 4-vector $\lambda^{sr,\mu}$ with spin index and 4-vector $\varrho^\mu$ as
\begin{eqnarray}
\lambda^{sr,\mu}&=&(\lambda^{sr,0},\pmb{\lambda}^{sr})=\left(\delta^{s,r},s \delta^{s,r} {\bf n}_3 + is  \delta^{-s,r} {\bf n}_2
+ \delta^{-s,r} {\bf n}_1 \right),\nonumber\\
\varrho^\mu &=&\frac{1}{2}\left(  \sqrt{E+p} + \sqrt{E-p}, (\sqrt{E+p} - \sqrt{E-p})\hat{\bf p}\right)
\end{eqnarray}
We can rewrite $\lambda^{sr,\mu}$ in the matrix form
\begin{eqnarray}
\lambda^{\mu}&=&\left(1,{\bf n}_3 \sigma_3^T + {\bf n}_2\sigma_2^T
+  {\bf n}_1 \sigma_1^T \right).
\end{eqnarray}
where the superscript $T$  denotes the transpose of a matrix.
Here we have introduced  two transverse unit 3-vector ${\bf n}_1$ and ${\bf n}_2$ orthogonal to ${\bf n}_3$
\begin{eqnarray}
{\bf n}_2 &=& \frac{\hat{ \bf z}\times {\bf n}_3}{|\hat{ \bf z}\times {\bf n}_3|}
=\left(-\sin\varphi, \cos\varphi, 0 \right),\\
{\bf n}_1 &=&{\bf n}_2\times {\bf n}_3
=  \left( \cos\vartheta \cos\varphi, \cos\vartheta \sin\varphi, -\sin\vartheta\right)
\end{eqnarray}
where $\hat{\bf z}$ is the unit vector along the $z$-axis. When we set ${\bf n}_3=\hat{\bf z}$, we will find that ${\bf n}_2=\hat{\bf y}$ and   ${\bf n}_1=\hat{\bf x}$. The first and second terms in Eq.(\ref{Lambda-sr-dirac}) can be
further dealt with by using the following identities
\begin{eqnarray}
\label{spinor-eq-1}
{u^{r\dagger}({\bf p})} u^s({\bf p})=2E \lambda^{sr,0},\ \ \
{u^{r\dagger}({\bf p})}{\pmb{\Sigma}} u^s({\bf p})
= 2 m \pmb{\lambda}^{sr} + 4 (\pmb{\varrho}\cdot\pmb{\lambda}^{rs})\pmb{\varrho}
\end{eqnarray}
However, in order  to deal with the last term including the derivative with the momentum on the Dirac spinors  in Eq.(\ref{Lambda-sr-dirac}),  we need to know whether the two-component $\xi^s$ depends on the momentum or not.

\subsection{Polarization along the fixed direction}
If the spin quantization direction ${\bf n}_3$  does not depend on the momentum ${\bf p}$, then the derivative does not act on the two-component spinor $\xi^s$. Using the identities
\begin{eqnarray}
{\pmb{ \nabla}}_p\sqrt{p\cdot\sigma}
&=&\frac{1}{2E}\pmb{\varrho} - \frac{1}{2E} \varrho_0\hat{\bf p}(\hat{\bf p} \cdot\pmb{\sigma})
-\frac{1}{p} (\pmb{\varrho}\cdot \hat{\bf p})\left[\pmb{\sigma}-\hat{\bf p}(\hat{\bf p} \cdot\pmb{\sigma})  \right]\\
{\pmb{ \nabla}}_p\sqrt{p\cdot\bar\sigma}
&=&\frac{1}{2E}\pmb{\varrho} + \frac{1}{2E} \varrho_0\hat{\bf p}(\hat{\bf p} \cdot\pmb{\sigma})
+\frac{1}{p} (\pmb{\varrho}\cdot \hat{\bf p})\left[\pmb{\sigma}-\hat{\bf p}(\hat{\bf p} \cdot\pmb{\sigma})  \right]
\end{eqnarray}
we have
\begin{eqnarray}
\label{spinor-eq-2}
\left[ {\pmb{ \nabla}}_p {u^{r\dagger}({\bf p})} \right]  u^s({\bf p}) &=&
\frac{2}{E}\varrho_0\lambda_0^{sr}\pmb{\varrho}
+\frac{2}{p}i(\hat{\bf p}\cdot \pmb{\varrho})\pmb{\varrho} \times \pmb{\lambda}^{sr}
\end{eqnarray}
Substituting  this equation and  Eq.(\ref{spinor-eq-1}) into the operator $\Lambda^{sr}_{\bf p}$ in Eq.(\ref{Lambda-sr-dirac}),  we get
\begin{eqnarray}
\label{Lambda-Dirac}
\Lambda^{sr}_{\bf p}&=& \Lambda_{\bf p}\delta^{sr} + \frac{1}{2}\pmb{\Omega}_{\bf p} \cdot\pmb{\lambda}^{sr}
\end{eqnarray}
where $\Lambda_{\bf p}$ is given in Eq.(\ref{Lambda}) and $\pmb{\Omega}_{\bf p}$ is an effective vorticity vector
 and defined by
\begin{eqnarray}
\pmb{\Omega}_{\bf p}= \left(\pmb{\omega}-\frac{\pmb{\varepsilon}\times {\bf p}}
{E_{{\bf p}}+m}\right)
\end{eqnarray}
Then $\Lambda^{sr}_{{\bf k}-{\bf q}/2}$ in the distribution function (\ref{f-sr-Dirac}) is given by
\begin{eqnarray}
\Lambda^{sr}_{{\bf k}-{\bf q}/2}&=& \Lambda_{{\bf k}-{\bf q}/2}\delta^{sr}
+\frac{1}{2} \pmb{\Omega}_{{\bf k}-{\bf q}/2}^{\textrm{D}}
\cdot\pmb{\lambda}^{sr}
\end{eqnarray}
Using the similar expansion to the Eq.(\ref{Expansion-BE}) for fermions up to the first order
\begin{eqnarray}
\frac{1}{e^{X+Y} + 1}
&=&f_F(X)+f'_F(X)Y - \frac{1}{2}f''_F(X)C  +\cdots\hspace{1cm}
\end{eqnarray}
where
\begin{eqnarray}
X&=&b_0 E_{{\bf k}-{\bf q}/2} - {\bf b}\cdot({\bf k}-{\bf q}/2) - \alpha,\ \ \
Y = -\Lambda_{{\bf k}-{\bf q}/2} ,\\
C&=&[X,Y]=i\left[\left(b_0 \pmb{\varepsilon}+ \pmb{\omega}\times {\bf b}\right)
\cdot \left( {\bf k}-\frac{\bf q}{2}\right) - ({\bf b}\cdot\pmb{\varepsilon})E_{{\bf k}-{\bf q}/2}\right]
\end{eqnarray}
we can expand  the distribution function  up to the first order.
The zeroth-order result is just the Fermi-Dirac distribution
\begin{eqnarray}
f^{(0)}_{rs}(x,k)
&=&f_F(b\cdot p-\alpha)\delta^{sr}
\end{eqnarray}
and the first-order result is given by
\begin{eqnarray}
f^{(1)}_{rs}(x,k)
= f'_F(b\cdot p-\alpha)\left[E_{\bf k}\pmb{\varepsilon}\cdot{\bf x} - {\bf k}\cdot (\pmb{\varepsilon}t + \pmb{\omega}\times{\bf x})\right]\delta^{sr} -\frac{1}{2}f'_F(b\cdot p-\alpha)
\pmb{\Omega}_{\bf k}\cdot\pmb{\lambda}^{sr}
\end{eqnarray}
Similar to the distribution function for scalar field, the first term in $f^{(1)}_{rs}(x,k)$ above can be absorbed into the $f^{(0)}_{rs}(x,k)$ with the $b^\mu$ replaced by $\beta^\mu$.  After this rearrangement of the contribution, we obtain the final distribution function  up to the first order
\begin{eqnarray}
f_{rs}(x,k)
&=& f_F(\beta\cdot p-\alpha)\delta^{sr}
 -\frac{1}{2}f'_F(\beta\cdot p-\alpha)
 \pmb{\Omega}_{\bf k} \cdot\pmb{\lambda}^{sr}
\end{eqnarray}
or  in the matrix form
\begin{eqnarray}
f(x,k)
&=& f_F(\beta\cdot p-\alpha)\cdot 1
 -\frac{1}{2}f'_F(\beta\cdot p-\alpha)
 \pmb{\Omega}_{\bf k}
 \cdot\left({\bf n}_3 \sigma_3 +{\bf n}_2 \sigma_2 + {\bf n}_1 \sigma_1 \right)
\end{eqnarray}
Then the local polarization along the fixed direction ${\bf n}_3$ follows as
\begin{eqnarray}
P(x,k) &\equiv& \frac{f_{+,+}(x,k) - f_{-,-}(x,k)}
{f_{+,+}(x,k) + f_{-,-}(x,k)}
=-\frac{ f'_F}{2f_F }\pmb{\Omega}_{\bf k}\cdot{\bf n}_3
\end{eqnarray}
where we have suppressed the argument $\beta\cdot p-\alpha$ in the last expression for brevity.

Let us compare our result with the one in Eq.(\ref{P-x-k-0}). Three differences are obvious: Firstly,
we have no  $1/m$ term which becomes singular when the mass approaching zero. Secondly, we have no
 term proportional to $(\pmb{\omega}\cdot {\bf k}) {\bf k}$ in the middle term of Eq.(\ref{P-x-k-0}).
Thirdly, the contribution from acceleration term ${\pmb{\varepsilon}\times{\bf k}}$ is suppressed by $E_{\bf k}/(E_{\bf k}+m)$ relative to the term proportional to
$\pmb{\omega}$.

\subsection{Polarization along the momentum direction }

For the helicity polarization, the spin quantization direction ${\bf n}_3$ is along the particle's momentum
${\bf p}= p (\sin\theta \cos\phi, \sin\theta \sin\phi, \cos\theta)$ with $p=|{\bf p}|$,
we have $\vartheta=\theta, \varphi=\phi$
\begin{eqnarray}
\label{n-123-h}
{\bf n}_3 = \hat{\bf p}  = {\bf e}_{ p},\ \ \
{\bf n}_2 = \frac{\hat{ \bf z}\times \hat{\bf p}}{|\hat{ \bf z}\times \hat{\bf p}|} = {\bf e}_{ \phi},\ \ \
{\bf n}_1 = {\bf n}_2\times \hat{\bf p}={\bf e}_{ \theta}
\end{eqnarray}
and the helicity spinor
\begin{eqnarray}
\label{Xi-helicity}
\xi^{s}=(- i)^{\frac{1-s}{2}}\left(\begin{array}{c}
          \cos\frac{s\theta+(1-s)\pi/2}{2}e^{-i\frac{\phi+(1-s)\pi/2}{2}} \\
          \sin\frac{s\theta+(1-s)\pi/2}{2} e^{i\frac{\phi+(1-s)\pi/2}{2}}
        \end{array}\right)
\end{eqnarray}
In such case, the derivative in the last term of the Eq.(\ref{Lambda-sr-dirac}) does  act on the two-component spinor $\xi^r$. It is easy to verify that
\begin{eqnarray}
\label{d-spinor-h}
\pmb{\nabla}_p\xi^{r}&=&
-\frac{i{\bf e}_\phi}{2p\sin\theta} \sigma_3\xi^r + \frac{ {\bf e}_\theta}{2p} r \xi^{-r}
\end{eqnarray}
With this  contribution, we obtain extra term  $\delta\Lambda^{sr}_{{\bf p}\parallel}$ in comparison with  the $\Lambda^{sr}_{\bf p}$  for the fixed spin direction
\begin{eqnarray}
& &\left[{\pmb{\varepsilon}}\cdot{\bf K}+ {\pmb{\omega}}\cdot{\bf J}, a_{\bf p}^{s\dagger}\right]
= \left(\Lambda^{sr}_{\bf p}+\delta\Lambda^{sr}_{{\bf p}\parallel} \right){a^r_{\bf p}}^\dagger
\end{eqnarray}
where $\Lambda^{sr}_{\bf p}$ is given by Eq.(\ref{Lambda-Dirac})  with ${\bf n}_3, {\bf n}_2, {\bf n}_1$ designated  in Eq.(\ref{n-123-h}) and
the extra term is given by
\begin{eqnarray}
\delta\Lambda^{sr}_{{\bf p}\parallel}  \equiv
\frac{E_{\bf p}}{2p}\pmb{\mathcal{E}}_{\bf p}\cdot
\left(-{{\bf e}_\phi}\cot \theta s\delta^{rs} + {\bf e}_\phi \delta^{-r,s}
- i{\bf e}_\theta s\delta^{-r,s}\right)
\end{eqnarray}
Then the distribution function with helicity index is given by
\begin{eqnarray}
f_{rs}(x,k)&=&\int d^3 {\bf q} e^{-i(E_{{\bf k}+{\bf q}/2}-E_{{\bf k}-{\bf q}/2})t+i{\bf q}\cdot{\bf x}} \nonumber\\
& &\times\left[\frac{1}{e^{b_0 E_{{\bf k}-{\bf q}/2} - {\bf b}\cdot({\bf k}-{\bf q}/2) - \alpha - \Lambda_{{\bf k}-{\bf q}/2} -\delta\Lambda_{{\bf k}-{\bf q}/2\parallel}}+1}\right]^{sr}
\delta\left({\bf q}\right)
\end{eqnarray}
The extra term  contribute to the first-order distribution function
\begin{eqnarray}
f^{(1)}_{rs}(x,k)
&=& f'_F(b\cdot k-\alpha)\left[E_{\bf k}\pmb{\varepsilon}\cdot{\bf x}
- {\bf k}\cdot (\pmb{\varepsilon}t + \pmb{\omega}\times{\bf x})\right]\delta^{sr}
 -\frac{1}{2}f'_F(b\cdot k-\alpha)
 \pmb{\Omega}_{\bf k}\cdot\pmb{\lambda}^{sr}\nonumber\\
& &+\frac{1}{2}f'_F(b\cdot k-\alpha)\frac{E_{\bf k}}{k}\pmb{\mathcal{E}}_{\bf k}
\cdot \left({{\bf e}_\phi}\cot \theta  s\delta^{rs}
-{\bf e}_\phi \delta^{-r,s}
+i{\bf e}_\theta s\delta^{-r,s}\right)\ \ \ \
\end{eqnarray}
where the last term is additional contribution compared to the fixed spin direction.
After the rearrangement  from $b^\mu$ to $\beta^\mu$, we obtain the final distribution function  up to the first order
\begin{eqnarray}
f_{rs}(x,k)
&=& f_F(\beta\cdot p-\alpha)\delta^{sr}
 -\frac{1}{2}f'_F(\beta\cdot p-\alpha)
  \pmb{\Omega}_{\bf k}\cdot\pmb{\lambda}^{sr}\nonumber\\
& &+\frac{1}{2}f'_F(\beta\cdot k-\alpha)\frac{E_{\bf k}}{k}\pmb{\mathcal{E}}_{\bf k}
\cdot \left({{\bf e}_\phi}\cot \theta  s\delta^{rs}
-{\bf e}_\phi \delta^{-r,s}
+i{\bf e}_\theta s\delta^{-r,s}\right)
\end{eqnarray}
or in matrix form
\begin{eqnarray}
f(x,k)
&=& f_F(\beta\cdot p-\alpha)\cdot 1
 -\frac{1}{2}f'_F(\beta\cdot p-\alpha)
  \pmb{\Omega}_{\bf k}
 \cdot \left({\bf n}_3 \sigma_3 +{\bf n}_2 \sigma_2 + {\bf n}_1 \sigma_1 \right)\nonumber\\
& &+\frac{1}{2}f'_F(\beta\cdot k-\alpha)\frac{E_{\bf k}}{k}\pmb{\mathcal{E}}_{\bf k}
\cdot \left({{\bf e}_\phi}\cot \theta \sigma_3
-{\bf e}_\phi \sigma_1
+ {\bf e}_\theta\sigma_2\right)
\end{eqnarray}
The  helicity polarization with ${\bf n}_3={\bf e}_p$ is given by
\begin{eqnarray}
P(x,k) &=&-\frac{ f'_F}{2f_F } \pmb{\Omega}_{\bf k}\cdot{\bf n}_3
+\frac{f'_F}{2f_F} \frac{E_{\bf k}}{k}\cot\theta\,
\pmb{\mathcal{E}}_{\bf k} \cdot {{\bf e}_\phi}
\end{eqnarray}
The last term will contribute to additional helicity polarization which is absent in the formalism in Eq.(\ref{P-x-k-0}).

\subsection{Polarization perpendicular to the momentum}
It is also interesting to consider the polarization perpendicular to the momentum. We have two independent directions perpendicular to the momentum.
One choice of  the spin quantization  ${\bf n}_3$ is  along the direction ${\bf e}_\phi $. This choice can be  fulfilled by  $\vartheta=\pi/2, \varphi=\phi+\pi/2$ and obtain
\begin{eqnarray}
{\bf n}_3=  {\bf e}_\phi,\ \ \
{\bf n}_2 = \hat{ \bf k}\times {\bf n}_3 ,\ \ \
{\bf n}_1 = {\bf n}_2\times {\bf n}_3= - \hat{\bf k}
\end{eqnarray}
Substituting it into Eq.(\ref{Lambda-sr-dirac}) and using the  relation for this specific case
\begin{eqnarray}
\pmb{\nabla}_p\xi^{r}&=&
-\frac{i{\bf e}_\phi}{2p\sin\theta} \sigma_3\xi^r
\end{eqnarray}
we  obtain  extra term  $\delta\Lambda^{sr}_{{\bf p}\phi}$ to  the $\Lambda^{sr}_{\bf p}$ for the fixed spin direction given in
(\ref{Lambda-Dirac})
\begin{eqnarray}
& &\left[{\pmb{\varepsilon}}\cdot{\bf K}+ {\pmb{\omega}}\cdot{\bf J}, a_{\bf p}^{s\dagger}\right]
= \left(\Lambda^{sr}_{\bf p}+\delta\Lambda^{sr}_{{\bf p}\phi} \right){a^r_{\bf p}}^\dagger
\end{eqnarray}
where
\begin{eqnarray}
\delta\Lambda^{sr}_{\bf{p}\phi}  &=&
\frac{E_{\bf p}}{2p} \pmb{\mathcal{E}}_{\bf p}\cdot
\left(- {i {\bf e}_\phi }\cot\theta s\delta^{r,-s} + {{\bf e}_\phi} \delta^{-r,s}\right)
\end{eqnarray}
Substituting this contribution into the distribution function
\begin{eqnarray}
f_{rs}(x,k)
&=&\int d^3 {\bf q} e^{-i(E_{{\bf k}+{\bf q}/2}-E_{{\bf k}-{\bf q}/2})t} e^{i{\bf q}\cdot{\bf x}}
\left[\frac{1}{e^{b_0 E_{{\bf k}-{\bf q}/2} - {\bf b}\cdot({\bf k}-{\bf q}/2) - \alpha - \Lambda_{\bf p}
-\delta\Lambda_{{\bf p}\phi}}+1}\right]^{sr}
\delta\left({\bf q}\right),
\end{eqnarray}
we obtain the  final distribution function  up to the first order
\begin{eqnarray}
f_{rs}(x,k)
&=& f_F(\beta\cdot p-\alpha)\delta^{sr}
 -\frac{1}{2}f'_F(\beta\cdot p-\alpha)
  \pmb{\Omega}_{\bf k} \cdot {\pmb{\lambda}}^{sr}\nonumber\\
& &+\frac{1}{2}f'_F(\beta\cdot k-\alpha)\frac{E_{\bf k}}{k}\pmb{\mathcal{E}}_{\bf k}
\cdot \left( i {{\bf e}_\phi}\cot \theta  s\delta^{r,-s}
-{\bf e}_\phi \delta^{-r,s}\right)
\end{eqnarray}
or in matrix form
\begin{eqnarray}
f(x,k)
&=& f_F(\beta\cdot p-\alpha)\cdot 1
 -\frac{1}{2}f'_F(\beta\cdot p-\alpha)
 \pmb{\Omega}_{\bf k}
 \cdot \left({\bf n}_3 \sigma_3 +{\bf n}_2 \sigma_2 + {\bf n}_1 \sigma_1 \right)\nonumber\\
& &+\frac{1}{2}f'_F(\beta\cdot k-\alpha)\frac{E_{\bf k}}{k}\pmb{\mathcal{E}}_{\bf k}
\cdot \left({{\bf e}_\phi}\cot \theta  \sigma_3
-{\bf e}_\phi \sigma_1 \right)
\end{eqnarray}
The   polarization can be given by
\begin{eqnarray}
P(x,k) =-\frac{ f'_0}{2f_0 }\pmb{\Omega}_{\bf k}\cdot{\bf n}_3
\end{eqnarray}
We note that the polarization  receive no extra contribution except for designating  ${\bf n}_3$ as ${\bf e}_\phi$.

Another choice of the independent transverse direction is along the direction ${\bf e}_\theta$, which can be obtained by
$\vartheta =\theta+\pi/2, \varphi=\phi$. In such case, we have
\begin{eqnarray}
{\bf n}_3=  {\bf e}_\theta,\ \ \
{\bf n}_2 =  {\bf e}_\phi ,\ \ \
{\bf n}_1 = -{\bf e}_p
\end{eqnarray}
and the derivative on the spinor satisfies the same relation as the helicity polarization (\ref{d-spinor-h}).
Along the same procedure as in the cases ${\bf n}_3={\bf e}_p$ and ${\bf n}_3={\bf e}_\phi$ above, we have the relation
\begin{eqnarray}
& &\left[{\pmb{\varepsilon}}\cdot{\bf K}+ {\pmb{\omega}}\cdot{\bf J}, a_{\bf p}^{s\dagger}\right]
= \left(\Lambda^{sr}+\delta\Lambda^{sr}_{{\bf p}\theta} \right){a^r_{\bf p}}^\dagger
\end{eqnarray}
where
\begin{eqnarray}
\delta\Lambda^{sr}_{{\bf p}\theta}  \equiv
\frac{E_{\bf p}}{2p} \pmb{\mathcal{E}}_{\bf k} \cdot
\left( { {\bf e}_\phi}\cot\theta \delta^{r,-s} - {\bf e}_\phi s \delta^{r,s}
-{ i{\bf e}_\theta} s\delta^{-r,s} \right)
\end{eqnarray}
 which leads to the final  first-order result for the distribution function is
\begin{eqnarray}
f_{sr}(x,k)
&=& f_F(\beta\cdot p-\alpha)\delta^{sr}
 -\frac{1}{2}f'_F(\beta\cdot p-\alpha)
 \pmb{\Omega}_{\bf k} \cdot\pmb{\lambda}^{sr}\nonumber\\
& &+\frac{1}{2}f'_F(\beta\cdot k-\alpha)\frac{E_{\bf k}}{k} \pmb{\mathcal{E}}_{\bf k}
 \left(- {{\bf e}_\phi}\cot \theta  \delta^{r,-s}
+{\bf e}_\phi s \delta^{r,s}
+ i {\bf e}_\theta s\delta^{-r,s}\right)\hspace{1cm}
\end{eqnarray}
or in matrix form
\begin{eqnarray}
f(x,k)
&=& f_F(\beta\cdot p-\alpha)\cdot 1
 -\frac{1}{2}f'_F(\beta\cdot p-\alpha)
 \pmb{\Omega}_{\bf k}
 \cdot \left({\bf n}_3 \sigma_3 +{\bf n}_2 \sigma_2 + {\bf n}_1 \sigma_1 \right)\nonumber\\
& &+\frac{1}{2}f'_F(\beta\cdot k-\alpha)\frac{E_{\bf k}}{k} \pmb{\mathcal{E}}_{\bf k}
\cdot \left(-{{\bf e}_\phi}\cot \theta \sigma_1
+{\bf e}_\phi \sigma_3
+ {\bf e}_\theta\sigma_2\right)
\end{eqnarray}
The polarization can be given by
\begin{eqnarray}
P &=&-\frac{ f'_F}{2f_F } \pmb{\Omega}_{\bf k}\cdot{\bf n}_3
+\frac{f'_F}{2f_F }\frac{E_{\bf k}}{k} \pmb{\mathcal{E}}_{\bf k}
\cdot {{\bf e}_\phi}
\end{eqnarray}
We note that the last term is an extra contribution  to the polarization compared to the fixed spin direction.

\section{Vector field}
\label{sec:vector}

Now we proceed to the charge vector field with the Lagrangian density
\begin{eqnarray}
 \mathcal{L} &=& -\frac{1}{2} F^\dagger_{\mu\nu}F^{\mu\nu}   + m^2 A^\dagger_\mu A^\mu
\end{eqnarray}
where the field tensor $F^{\mu\nu}$ is defined as
\begin{eqnarray}
F_{\mu\nu} = \partial_\mu  A_\nu  - \partial_\nu  A_\mu
\end{eqnarray}
The Euler-Lagrangian  equation leads to the Proca equation
\begin{eqnarray}
\label{Proca-0-C}
\partial_\mu \partial^\mu A_\nu  + m^2 A_\nu = 0 ,\ \ \ \partial_\mu \partial^\mu A_\nu^\dagger  + m^2 A_\nu^\dagger = 0
\end{eqnarray}
with  the constraint condition
\begin{eqnarray}
\label{constraint-A-C}
 \partial^\nu A_\nu = 0.
\end{eqnarray}
to get rid of the spin-0 contribution.
The general solution can be expressed as a Fourier transform
\begin{eqnarray}
\label{Amu-C}
A_\mu (x) =  \int \frac{d^3{\bf p}}{(2\pi)^3}\frac{1}{\sqrt{2E_{\bf p}}}
\sum_{r=1}^3\left( a_{\bf p}^{r}\epsilon_\mu^{r}({\bf p}) e^{-ip\cdot x}
+ b_{\bf p}^{r\dagger}\eta_\mu^{r*}({\bf p}) e^{ip\cdot x}\right)
\end{eqnarray}
with the constraint from (\ref{constraint-A-C})
\begin{eqnarray}
p^\mu \epsilon_\mu^{r}({\bf p}) = 0,\ \ \ p^\mu \eta_\mu^{r}({\bf p}) = 0
\end{eqnarray}
and  the expression
\begin{eqnarray}
\epsilon^{r\mu}({\bf p})=\left(\frac{{\bf p}\cdot{\bf n}_r}{m}, {\bf n}_r
+\frac{{\bf p}\cdot{\bf n}_r}{m(E+m)}{\bf p}\right).
\end{eqnarray}
Here  ${\bf n}_r$ ($r=1,2,3$) is real orthogonal unit vector satisfying ${\bf n}_3 = {\bf n}_1\times {\bf n}_2$.
For our choice of $\epsilon_\mu^{r}({\bf p})$, we have $\epsilon^{r*}_\mu({\bf p}) = \epsilon^{r}_\mu({\bf p})$.
The antiparticle part $\eta_\mu^{r}({\bf p})$ can be obtained in the same way as $\epsilon_\mu^{r}({\bf p})$.
As usual, we choose ${\bf n}_3$ as the spin quantization direction.
The creation and annihilation operators obey the commutation rules
\begin{eqnarray}
\left[ a^r_{\bf p},  a_{\bar{\bf p}}^{s\dagger}\right]=(2\pi)^3 \delta^3({\bf p}-\bar{\bf p})\delta^{rs},\ \ \
\left[ b^r_{\bf p},  b_{\bar{\bf p}}^{s\dagger}\right]=(2\pi)^3 \delta^3({\bf p}-\bar{\bf p})\delta^{rs}.
\end{eqnarray}
The conserved charge current of the Proca theory is given by
\begin{eqnarray}
\label{j-Proco}
j^{\mu} &=& -i\left( \partial^\mu  A_\nu^\dagger A^\nu - A^\dagger_\nu \partial^\mu A^\nu\right),
\end{eqnarray}
The canonical energy-momentum tensor of the Proca theory is
\begin{eqnarray}
T^{\mu\nu}
&=& \frac{1}{2}g^{\mu\nu}F^\dagger_{\alpha\beta}F^{\alpha\beta} - F^{\mu\alpha\dagger}\partial^\nu A_\alpha
- \partial^\nu A^\dagger_\alpha F^{\mu\alpha}
-m^2 g^{\mu\nu}A^\dagger_\alpha A^\alpha
\end{eqnarray}
The angular momentum density is then obtained
\begin{eqnarray}
\mathcal{M}^{\mu\alpha\beta}
&=&  x^\alpha T^{\mu\beta} - x^\beta T^{\mu\alpha}
 +\left( A^{\alpha\dagger} F^{\mu\beta} -  A^{\beta\dagger} F^{\mu\alpha}\right)
 +\left( F^{\mu\beta\dagger} A^{\alpha} -   F^{\mu\alpha\dagger} A^{\beta}\right)
\end{eqnarray}
The  conserved  charge and energy-momentum can be expressed as
\begin{eqnarray}
Q = \int\frac{d^3{\bf p}}{(2\pi)^3}
\sum_{s=1}^3 \left( a_{\bf p}^{s\dagger}   a_{\bf p}^{s} -b_{\bf p}^{s}   b_{\bf p}^{s\dagger} \right),\ \ \
P^\mu = \int\frac{d^3{\bf p}}{(2\pi)^3}
\sum_{s=1}^3 p^{\mu}\left( a_{\bf p}^{s\dagger}   a_{\bf p}^{s}
+b_{\bf p}^{s}   b_{\bf p}^{s\dagger} \right)
\end{eqnarray}
The angular momentum tensor can be given by
\begin{eqnarray}
K^{i}
&=& \int \frac{d^3{\bf p}}{(2\pi)^3}
 \sum_{s=1}^3\sum_{r=1}^3\nonumber\\
& &\left\{
 \sqrt{E_{\bf p}} \epsilon_\alpha^{s*}({\bf p})a_{\bf p}^{s\dagger} i \partial_i^p
 \left[ \sqrt{E_{\bf p}} \epsilon^{r\alpha }({\bf p})a_{\bf p}^{r}\right]
 - \sqrt{E_{\bf p}}  \eta_\alpha^{s}({\bf p}) b_{\bf p}^{s}  i\partial_i^p
\left[\sqrt{E_{\bf p}} \eta^{r \alpha *}({\bf p})b_{\bf p}^{r\dagger}\right]\frac{}{}\right.\nonumber\\
& &\left.\ \ \
-i a_{\bf p}^{s\dagger}a_{\bf p}^{r}
\left[ \epsilon^{s0*}({\bf p})  \epsilon^{r i }({\bf p})
-  \epsilon^{s i*}({\bf p}) \epsilon^{r 0}({\bf p})  \right]\right.\nonumber\\
& &\left.\ \ \   +i b_{\bf p}^{s} b_{\bf p}^{r\dagger}
\left[ \eta^{s 0}({\bf p}) \eta^{r i *}({\bf p})
- \eta^{s i}({\bf p}) \eta^{r 0 *}({\bf q})   \right]\frac{}{}\right\},\\
J^{i}
&=&\epsilon^{0ijk} \int\frac{d^3{\bf p}}{(2\pi)^3}
\sum_{s=1}^3\sum_{r=1}^3\nonumber\\
& &\left\{\frac{}{}
- a_{\bf p}^{s\dagger} \epsilon^{s \alpha *}({\bf p})
p^k i\partial^p_j\left[ \epsilon^{r }_\alpha({\bf p})  a_{\bf p}^{r}\right]
 -  b_{\bf p}^{s}  \eta^{s\alpha }({\bf p})
p^k i\partial_j^p \left[\eta^{r*}_\alpha({\bf p})b_{\bf p}^{r\dagger}\right]
\frac{}{}\right.\nonumber\\
& &\hspace{9pt} \left.-i a_{\bf p}^{s\dagger}  a_{\bf p}^{r}
 \epsilon^{s j *}({\bf p}) \epsilon^{r k}({\bf p})
+i b_{\bf p}^{s}  b_{\bf p}^{r\dagger}
\eta^{s j}({\bf p})\eta^{r k* }({\bf p})
\frac{}{}\right\}
\end{eqnarray}
It is straightforward to  obtain the commutation relations
\begin{eqnarray}
\left[Q, a_{\bf p}^{s\dagger}\right] =  a_{{\bf p}}^{s\dagger},\ \ \
\left[b_\mu P^\mu, a_{\bf p}^{s\dagger}\right] = b_\mu p^\mu a_{{\bf p}}^{s\dagger},\ \ \
\left[{\pmb{\varepsilon}}\cdot{\bf K} + {\pmb{\omega}}\cdot{\bf J}, a_{\bf p}^{s\dagger}\right]
=\sum_r \Lambda_{\bf p}^{sr}{a_{\bf p}}^{r\dagger}
\end{eqnarray}
where the operator $\Lambda_{\bf p}^{sr}$ for the vector field is defined by
\begin{eqnarray}
\Lambda_{\bf p}^{sr}&=&- \,\epsilon^{s}_\alpha({\bf p}) \epsilon^{r \alpha*}({\bf p})
{E_{\bf p}} \pmb{\mathcal{E}}_{\bf p} \cdot i\pmb{\nabla}_{p}
-\frac{i}{2E_{\bf p}} \,\epsilon^{s}_\alpha({\bf p})\epsilon^{r \alpha*}({\bf p})
\pmb{\varepsilon}\cdot {\bf p}
\nonumber\\
& &- \,\epsilon^{s}_\alpha({\bf p})\left[i\pmb{\nabla}_{p}
\epsilon^{r \alpha*}({\bf p})a_{\bf p}^{r\dagger} \right]
\cdot {E_{\bf p}} \pmb{\mathcal{E}}_{\bf p}
 -i \pmb{\varepsilon}\cdot
\left[ \epsilon^{r0*}({\bf p}) \pmb{ \epsilon}^{s  }({\bf p})
-   \epsilon^{s 0}({\bf p})\pmb{ \epsilon}^{r*}({\bf p})  \right]\nonumber\\
 & & -i\pmb{\omega}\cdot \left[ \pmb{\epsilon}^{r*}({\bf p})\times \pmb{ \epsilon}^{s  }({\bf p})
   \right]
\end{eqnarray}
It follow that the distribution function is given by
\begin{eqnarray}
\label{f-sr-vector}
& &f_{rs}(x,k)\equiv
\frac{1}{(2\pi)^3}\int d^3 {\bf q} e^{-i(E_{{\bf k}+{\bf q}/2}-E_{{\bf k}-{\bf q}/2})t+i{\bf q}\cdot{\bf x}}
\langle a^{s\dagger}_{{\bf k}-{\bf q}/{2}} a^r_{ {\bf k}+{\bf q}/{2}}\rangle_0\nonumber\\
&=&\int d^3 {\bf q} e^{-i(E_{{\bf k}+{\bf q}/2}-E_{{\bf k}-{\bf q}/2})t+i{\bf q}\cdot{\bf x}}
\left[\frac{1}{e^{b_0 E_{{\bf k}-{\bf q}/2} - {\bf b}\cdot({\bf k}-{\bf q}/2) - \alpha
- \Lambda_{{\bf k}-{\bf q}/2}}-1}\right]^{sr}
\delta\left({\bf q}\right)\hspace{1cm}
\end{eqnarray}
In the following, we will deal with this distribution function further by using  the  identities
\begin{eqnarray}
\label{relation-vector}
\epsilon_\mu^{r}({\bf p})\epsilon^{ s \mu*}({\bf p}) &=& -\delta^{r s}\nonumber\\
\epsilon^{s }_\alpha({\bf p})\partial^i_p \left[ \epsilon^{r\alpha *}({\bf p})\right]
&=&
- \frac{\epsilon^{ri*}({\bf p})}{E_{\bf p}+m}\epsilon^{s 0 }({\bf p})
+\frac{\epsilon^{r0*}({\bf p})}{E_{\bf p}+m}\epsilon^{s i }({\bf p})
{-{\bf n}_s\cdot \partial^i_p {\bf n}^*_r }
\end{eqnarray}

\subsection{Polarization along the fixed direction}

If the unit vectors ${\bf n}_1$, ${\bf n}_2$,  ${\bf n}_3$ are independent on the momentum,
the last term in Eq.(\ref{relation-vector}) will vanish and we obtain

\begin{eqnarray}
\label{Lambda-V-n}
\Lambda^{sr}_{\bf p}
&=& \Lambda_{\bf p}\delta^{sr} + i\pmb{\Omega}_{\bf p}
\cdot \left( {\bf n}_s \times {\bf n}_r\right)
\end{eqnarray}
Following the same line for the scalar field and the Dirac field and
using the expansion in Eq.(\ref{Expansion-BE}), we obtain the zeroth-order result for the vector particle
\begin{eqnarray}
f^{(0)}_{rs}(x,k)&=&f_B(b\cdot k-\alpha)\delta^{sr}
\end{eqnarray}
The first-order result is given by
\begin{eqnarray}
f^{(1)}_{rs}(x,k)
&=& f'_B(b\cdot k-\alpha)\left[E_{\bf k}\pmb{\varepsilon}\cdot{\bf x}
 - {\bf k}\cdot (\pmb{\varepsilon}t + \pmb{\omega}\times{\bf x})\right]
\delta^{sr}\nonumber\\
& & - i f'_B(b\cdot k-\alpha)
\pmb{\Omega}_{{\bf k}} \cdot
\left( {\bf n}_s \times {\bf n}_r\right)
\end{eqnarray}
As we all know, the  vector $\epsilon_\mu^{r}({\bf p})$ with $r=1,2,3$  is the linear polarization vector and does not correspond
to the spin eigenstate. The spin eigenstate  can be achieved by introducing circular  polarization operators
\begin{eqnarray}
a^{+}_{\bf p} =- \frac{1}{\sqrt{2}}\left(a^{1}_{\bf p}-ia^{2}_{\bf p} \right),\ \ \
a^{-}_{\bf p} = \frac{1}{\sqrt{2}}\left(a^{1}_{\bf p}+ia^{2}_{\bf p} \right),\ \ \
a^{0}_{\bf p} = a^{3}_{\bf p}
\end{eqnarray}
where $0$ and $\pm$  denote the spin components with $0$, $\pm 1$, respectively. With circular polarization indices, the vector field
can be given by
\begin{eqnarray}
\label{Amu-C-c}
A_\mu (x) =  \int \frac{d^3{\bf p}}{(2\pi)^3}\frac{1}{\sqrt{2E_{\bf p}}}
\sum_{r=0,\pm}\left( a_{\bf p}^{r}\epsilon_\mu^{r}({\bf p}) e^{-ip\cdot x}
+ b_{\bf p}^{r\dagger}\eta_\mu^{r*}({\bf p}) e^{ip\cdot x}\right)
\end{eqnarray}
where the polarization four-vectors with circular polarization index  is defined by
\begin{eqnarray}
\label{Amu-C-c}
\epsilon_\mu^{+}({\bf p})=-\frac{1}{\sqrt{2}}\left[\epsilon_\mu^{1}({\bf p})+ i\epsilon_\mu^{2}({\bf p}) \right], \ \
\epsilon_\mu^{-}({\bf p})=\frac{1}{\sqrt{2}}\left[\epsilon_\mu^{1}({\bf p})- i\epsilon_\mu^{2}({\bf p}) \right],\ \
\epsilon_\mu^{0}({\bf p})=\epsilon_\mu^{3}({\bf p})
\end{eqnarray}
We will use the same indices $r,s$ to denote linear or circular polarization.    With the circular polarization indices $0,\pm$, we have
\begin{eqnarray}
f^{(1)}_{rs}(x,p)
&=& f'_B(b\cdot k-\alpha)\left[E_{\bf k}\pmb{\varepsilon}\cdot{\bf x}
 - {\bf k}\cdot (\pmb{\varepsilon}t + \pmb{\omega}\times{\bf x})\right]
\delta^{sr}\nonumber\\
& & - i f'_B(b\cdot k-\alpha)
\pmb{\Omega}_{{\bf k}}
\cdot \left( {\bf n}_s \times {\bf n}_r^*\right)
\end{eqnarray}
where polarizaiton three-vector with circular indices is given by
\begin{eqnarray}
\label{Amu-C-c-n}
{\bf n}_+=-\frac{1}{\sqrt{2}}\left({\bf n}_1 + i {\bf n}_2 \right), \ \
{\bf n}_-=\frac{1}{\sqrt{2}}\left({\bf n}_1 - i {\bf n}_2 \right),\ \
{\bf n}_0={\bf n}_3
\end{eqnarray}
In order to obtain the non-trivial contribution for the spin alignment, we need the second-order result for the diagonal components
with linear polarization indices
\begin{eqnarray}
f_{rr}^{(2)}(x,k) &=&f^{(2)}(x,k)+\frac{1}{2}f''_B(b\cdot k-\alpha)\left( \pmb{\Omega}_{{\bf k}}\right)  ^2
-\frac{1}{2}f''_B(b\cdot k-\alpha)\left( \pmb{\Omega}_{{\bf k}}  \cdot {\bf n}_r\right)^2
\end{eqnarray}
where $f^{(2)}$ denotes the second-order contribution to the distribution function for scalar field as given in (\ref{f-2-KG}).
Summing $f^{(0)}_{rs}$ and $f^{(1)}_{rs}$ for non-diagonal components  and replacing $b^\mu$ with $\beta^\mu$, we obtain the spin  distribution function  up to the first order
\begin{eqnarray}
\label{f-rs-V-1}
f_{rs}(x,k)
&=& f_B(\beta\cdot k-\alpha)\delta^{sr}- i f'_B(\beta\cdot k-\alpha)
 \pmb{\Omega}_{{\bf k}}  \cdot \left( {\bf n}_s
\times {\bf n}_r^*\right)
\end{eqnarray}
This expression is valid for both  linear polarization indices $r,s=1,2,3$ and  circular polarization indices $r,s=0,\pm$.
For the linear polarization $r,s=1,2,3$, we have ${\bf n}_r^*={\bf n}_r $.
Summing $f^{(0)}_{rr}$, $f^{(1)}_{rr}$ and $f^{(2)}_{rr}$ for diagonal components with linear polarization indices and replacing $b^\mu$ with $\beta^\mu$, we obtain the diagonal distribution function  up to the second order
\begin{eqnarray}
f_{rr}(x,k) &=&f(x,k) +\frac{1}{2}f''_B(\beta\cdot k-\alpha) \left( \pmb{\Omega}_{{\bf k}} \right)^2
-\frac{1}{2}f''_B(\beta\cdot k-\alpha)\left(\pmb{\Omega}_{{\bf k}}  \cdot {\bf n}_r\right)^2
\end{eqnarray}
where $f(x,k)$ is  the distribution function for scalar field as given in (\ref{f-scalar}). It should be noted that this second-order expression
hold only for linear polarization indices $r,s=1,2,3$.

As we all know, the spin density matrix for the vector particles such as
$\phi$ and $K^{*0}$ mesons can be measured by their two-body decay channels
$\phi\rightarrow KK$ and $K^{*0}\rightarrow K\pi$, in which the distribution
of the decay products is related to the elements of spin density matrix by
\begin{eqnarray}
\label{Meson-decay}
\frac{dN}{d\cos\theta^*d\phi^*} &=&\frac{3}{8\pi}
\left[ 1 - \rho_{00}+(3\rho_{00}-1)\cos^2\theta^* \right.\nonumber\\
& & -\sqrt{2}\textrm{Re}\left( \rho_{+0} - \rho_{0-} \right)\sin\theta^*\cos\phi^*
+\sqrt{2}\textrm{Im} \left(\rho_{+0} - \rho_{0-} \right)\sin\theta^*\sin\phi^*\nonumber\\
& &\left. -2\textrm{Re} \rho_{+-} \sin^2\theta^*\cos(2\phi^*)
-2\textrm{Im} \rho_{+-} \sin^2\theta^*\sin(2\phi^*)\right]
\end{eqnarray}
where $\theta^*$ and $\phi^*$ are the polar and azimuthal angles of the
momentum of one final meson in the  rest frame of the initial vector mesons.
When we approximate the spin distribution function up to the first order
as given in Eq.(\ref{f-rs-V-1}), all the non-diagonal elements vanish and the diagonal element $\rho_{00}$ is $1/3$ which means that there is
no spin alignment.  At the second order, the spin alignment will receive nonzero  contribution because the local spin alignment can be given by
\begin{eqnarray}
\rho_{00}(x,k) &=& \frac{f_{00}}{f_{++}+f_{--}+f_{00}}
=\frac{1}{3} -\frac{1}{6}\frac{f_B''}{f_B}
\left[ \left( \pmb{\Omega}_{{\bf k}} \cdot {\bf n}_3\right)^2
-\frac{1}{3}\left(\pmb{\Omega}_{{\bf k}}\right)^2 \right].
\end{eqnarray}
We note that whether the spin alignment is less or greater than $1/3$ depends on the balance between the contribution
$\left( \pmb{\Omega}_{{\bf k}} \cdot {\bf n}_3\right)^2$ and $\left(\pmb{\Omega}_{{\bf k}}\right)^2/3$.

\subsection{Polarization along the momentum direction}
\label{sub:Polarization-p}
Now let us consider the helicity polarization with $\hat{\bf p}$ as the spin quantization direction
\begin{eqnarray}
\label{n-helicity}
{\bf n}_3 = {\bf e}_p,\  \ {\bf n}_2={\bf e}_\phi, \ \ {\bf n}_1={\bf e}_\theta
\end{eqnarray}
To calculate the last term in the second identity in Eq.(\ref{relation-vector}), we need the following relations
\begin{eqnarray}
\label{dn-V}
\frac{\partial}{\partial p}{\bf n}_s=0,\ \ \
\frac{\partial}{\partial\phi}{\bf n}_s=\hat{\bf z}\times {\bf n}_s,\ \ \
\frac{\partial}{\partial\theta}{\bf n}_s =  {\bf e}_\phi\times {\bf n}_s
\end{eqnarray}
With these relations,some additional terms will contribute the commutation relation
\begin{eqnarray}
\left[{\pmb{\varepsilon}}\cdot{\bf K} + {\pmb{\omega}}\cdot{\bf J}, a_{\bf p}^{s\dagger}\right]
&\equiv&\left(\Lambda^{sr}_{\bf p}{ +\delta \Lambda^{sr}_{{\bf p}}} \right){a^r_{\bf p}}^\dagger
\end{eqnarray}
where $\Lambda^{sr}_{\bf p}$ is given by Eq.(\ref{Lambda-V-n}) and the additional term $\delta\Lambda^{sr}_{{\bf p}\parallel}$ is
given by
\begin{eqnarray}
\delta\Lambda^{sr}_{{\bf p}}
&=&- \frac{iE_{\bf p} }{p}\left( \pmb{\mathcal{E}}_{\bf p}\cdot {\bf e}_\theta\right)
 {\bf e}_\phi \cdot\left({\bf n}_s\times {\bf n}_r\right)
- \frac{iE_{\bf p} }{p\sin\theta}
\left( \pmb{\mathcal{E}}_{\bf p}\cdot {\bf e}_\phi\right) \hat{\bf z}
\cdot \left({\bf n}_s\times {\bf n}_r\right)
\end{eqnarray}
From the definition
\begin{eqnarray}
f_{rs}(x,k)
&=&\int d^3 {\bf q} e^{-i(E_{{\bf k}+{\bf q}/2}-E_{{\bf k}-{\bf q}/2})t+i{\bf q}\cdot{\bf x}} \nonumber\\
& &\times\left[\frac{1}{e^{b_0 E_{{\bf k}-{\bf q}/2} - {\bf b}\cdot({\bf k}-{\bf q}/2) - \alpha - \Lambda_{{\bf k}-{\bf q}/2}
-\delta\Lambda_{{\bf k}-{\bf q}/2,\parallel}}-1}\right]^{sr}
\delta\left({\bf q}\right)\hspace{1cm}
\end{eqnarray}
 we obtain the distribution function up to the first order
\begin{eqnarray}
f_{rs}
&=& f_B
 - i f'_B \left[\pmb{\Omega}_{\bf k}
 -\frac{E_{\bf k} }{k}\left( \pmb{\mathcal{E}}_{\bf k} \cdot {\bf e}_\theta\right) {\bf e}_\phi
 -\frac{E_{\bf k}}{k\sin\theta}\left( \pmb{\mathcal{E}}_{\bf k}  \cdot {\bf e}_\phi \right) \hat{\bf z}\right]\cdot
\left( {\bf n}_s \times {\bf n}_r^*\right).
\end{eqnarray}
from which we find all the  non-diagonal element of the spin density matrix vanish and there is no spin alignment.
In order to obtain nonzero spin alignment, we need the second-order result for the diagonal elements
\begin{eqnarray}
f_{rr} &=&f_B+\frac{1}{2}f''_B \left[\pmb{\Omega}_{\bf k}
 -\frac{E_{\bf k} }{k}\left( \pmb{\mathcal{E}}_{\bf k} \cdot {\bf e}_\theta\right) {\bf e}_\phi
 -\frac{E_{\bf k}}{k\sin\theta}\left( \pmb{\mathcal{E}}_{\bf k}  \cdot {\bf e}_\phi \right) \hat{\bf z}\right]^2\nonumber\\
& &\hspace{0.5cm}-\frac{1}{2}f''_B\left\{\left[\pmb{\Omega}_{\bf k}
 -\frac{E_{\bf k} }{k}\left( \pmb{\mathcal{E}}_{\bf k} \cdot {\bf e}_\theta\right) {\bf e}_\phi
 -\frac{E_{\bf k}}{k\sin\theta}\left( \pmb{\mathcal{E}}_{\bf k}  \cdot {\bf e}_\phi \right) \hat{\bf z}\right]\cdot {\bf n}_r\right\}^2
 \hspace{1cm}
\end{eqnarray}
It follow that the spin alignment is given by
\begin{eqnarray}
\label{rho-00-h}
\rho_{00}
&=& \frac{1}{3} -\frac{1}{6}\cdot \frac{f_B''}{f_B}
\left\{\left[\left(\pmb{\Omega}_{\bf k}
 -\frac{E_{\bf k} }{k}\left( \pmb{\mathcal{E}}_{\bf k} \cdot {\bf e}_\theta\right) {\bf e}_\phi
 -\frac{E_{\bf k}}{k\sin\theta}\left( \pmb{\mathcal{E}}_{\bf k}  \cdot {\bf e}_\phi \right) \hat{\bf z}\right)\cdot {\bf n}_3\right]^2\right.\nonumber\\
& &\left.\hspace{2.1cm}-\frac{1}{3}\left[\pmb{\Omega}_{\bf k}
 -\frac{E_{\bf k} }{k}\left( \pmb{\mathcal{E}}_{\bf k} \cdot {\bf e}_\theta\right) {\bf e}_\phi
 -\frac{E_{\bf k}}{k\sin\theta}\left( \pmb{\mathcal{E}}_{\bf k}  \cdot {\bf e}_\phi \right) \hat{\bf z}\right]^2\right\}
\end{eqnarray}

\subsection{Polarization  perpendicular to the momentum}
Now let us consider the transverse polarization which is orthogonal to the momentum. Similar to the Dirac particle, we have two independent basis vectors. We can choose one group of basis vectors as
\begin{eqnarray}
\label{n-transverse-1}
{\bf n}_3 ={\bf e}_\phi ,\  \ {\bf n}_2={\bf e}_\theta, \ \ {\bf n}_1={\bf e}_p
\end{eqnarray}
and the other  group as
\begin{eqnarray}
\label{n-transverse-2}
{\bf n}_3 = {\bf e}_\theta,\  \ {\bf n}_2={\bf e}_p, \ \ {\bf n}_1={\bf e}_\phi
\end{eqnarray}
For both groups listed above, we follow the same routine as used for the helicity polarization and find that all the results  are the same as the ones given in the last subsection \ref{sub:Polarization-p} from Eq.(\ref{dn-V}) to Eq.(\ref{rho-00-h}). The only difference
is that we need replace ${\bf n}_1, {\bf n}_2, {\bf n}_3$ in Eq.(\ref{n-helicity}) by either Eq.(\ref{n-transverse-1}) or Eq.(\ref{n-transverse-2}).

\section{Summary}

\label{sec:summary}

We have revisited the spin polarization by thermal vorticity and proposed another formalism to calculate them directly from the spin-dependent
distribution function. We have calculated these spin-dependent distribution functions for spin-1/2 and spin-1 in global equilibrium with thermal vorticity. For  Dirac field with spin 1/2, the local polarization along the fixed direction ${\bf n}_3$ is given by
\begin{eqnarray}
P(x,k)=-\frac{ f'_F}{2f_F }\left(\pmb{\omega}-\frac{\pmb{\varepsilon}\times  {\bf k}}{E_{\bf k}+m}\right)\cdot{\bf n}_3,
\end{eqnarray}
For  vector field with spin 1, the spin alignment and all the diagonal elements of spin density matrix are absent up to the first order.  At the second order, the local spin alignment along  fixed direction ${\bf n}_3$ is given by
\begin{eqnarray}
\rho_{00}(x,k) = \frac{1}{3} -\frac{1}{6}\frac{f_B''}{f_B}
\left\{ \left[\left(\pmb{\omega}-\frac{\pmb{\varepsilon}\times  {\bf k}}{E_{\bf k}+m}\right)  \cdot {\bf n}_3\right]^2
-\frac{1}{3}\left(\pmb{\omega}-\frac{\pmb{\varepsilon}\times  {\bf k}}{E_{\bf k}+m}\right)^2 \right\},
\end{eqnarray}
Besides, we also find that when the spin quantization direction is dependent on the momentum, we will obtain additional contribution which is also different from the earlier prediction.  It will be valuable to make numerical simulation with these results and study the difference quantitatively
in the future.


\acknowledgments
We thank F. Becattini and S. Pu for helpful discussions. This work was supported in part by National Natural Science Foundation of China  under Nos. 11890710, 11890713, 12175123 and the Major Program of  Natural Science Foundation of Shandong Province under No. ZR2020ZD30.




\end{document}